\documentclass[12pt]{article}
\usepackage[margin=0.8in]{geometry} 
\usepackage{comment}
\usepackage{algorithm}
\usepackage{algpseudocode} 
\usepackage{multirow}
\usepackage{array}
\usepackage{amsfonts}
\usepackage{amssymb}
\usepackage[T1]{fontenc}
\usepackage{lmodern}  
\usepackage{graphicx}
\usepackage{float}
\usepackage{hyperref}
\usepackage{booktabs}
\usepackage{amsmath}
\usepackage{natbib}
\usepackage{bbm}
\usepackage{enumitem}
\usepackage{authblk} 

\title{\huge Adaptive clinical trial design with delayed treatment effects using elicited prior distributions}

\author[1]{\small James A. Salsbury\thanks{Corresponding author: jsalsbury1@sheffield.ac.uk}}
\author[1]{Jeremy E. Oakley}
\author[2]{Steven A. Julious}
\author[3]{Lisa V. Hampson}

\affil[1]{The School of Mathematical and Physical Sciences, The University of Sheffield, U.K.}
\affil[2]{The School of Medicine and Population Health, The University of Sheffield, U.K.}
\affil[3]{Advanced Methodology \& Data Science, Novartis Pharma AG, Basel, Switzerland}

\date{} 

\begin{document}

\maketitle

\begin{abstract}
\noindent Clinical trials with time-to-event endpoints, such as overall survival (OS) or progression-free survival (PFS), are fundamental for evaluating new treatments, particularly in immuno-oncology. However, modern therapies, such as immunotherapies and targeted treatments, often exhibit delayed effects that challenge traditional trial designs. These delayed effects violate the proportional hazards assumption, which underpins standard statistical methods like the Cox proportional hazards model and the log-rank test. Careful planning is essential to ensure trials are appropriately designed to account for the timing and magnitude of these effects. Without this planning, interim analyses may lead to premature trial termination if the treatment effect is underestimated early in the study. We present an adaptive trial design framework that incorporates prior distributions, elicited from experts, for delayed treatment effects. By addressing the uncertainty surrounding delayed treatment effects, our approach enhances trial efficiency and robustness, minimizing the risk of premature termination and improving the detection of treatment benefits over time. We present an example illustrating how interim analyses, informed by prior distributions, can guide early stopping decisions. To facilitate the implementation of our framework, we have developed free, open-source software that enables researchers to integrate prior distributions into trial planning and decision-making. This software provides a flexible, accessible tool for designing trials that more accurately evaluate modern therapies through adaptive trial designs.
\end{abstract}

\section{Introduction}

Clinical trials with time-to-event endpoints, such as overall survival (OS) or progression-free survival (PFS), are fundamental in evaluating new treatments in many therapeutic areas, especially in immuno-oncology. Randomised clinical trials (RCTs) remain the gold standard for such evaluations, but the advent of modern therapies, including immunotherapies and targeted treatments, has introduced complexities that challenge traditional trial designs. A significant issue arises when treatments exhibit delayed effects, where therapeutic benefits are not immediately apparent but only emerge after a certain period. Critically, key trial parameters---such as the length of the delay and the magnitude of the post-delay treatment effect---are inherently uncertain at the design stage, making it challenging to design trials that are both adequately powered and robust to this uncertainty. This motivates the use of principled elicitation methods to quantify and incorporate expert knowledge about these parameters directly into the trial design process.
Delayed treatment effects (DTEs) pose significant challenges in the design, analysis, and interpretation of clinical trials. A major issue is that they violate the proportional hazards (PH) assumption underlying the standard Cox proportional hazards model. Since the log-rank test is most powerful when the PH assumption holds, delayed effects can render these methods suboptimal or even inappropriate.\cite{Fine2007, Ristl2020, Mukhopadhyay2020, Sit2016, Freidlin2019, Magirr2021} Given that survival trials are typically event-driven and require lengthy follow-up, interim analyses are commonly incorporated to enable early evaluation and decision-making.\cite{Berry2012} However, when treatment effects are delayed, the risk of misinterpretation at interim is particularly acute---early data may underestimate the eventual benefit, potentially leading to premature termination for futility.\cite{Chen2013, Hoos2010, Menis2016, Xu2016, Korn2018} This is precisely the setting where a principled framework for futility decision-making, informed by prior knowledge about the delay, is most valuable.
In our previous work \cite{Salsbury2024}, we developed methods for eliciting key parameters---such as the length of delay and the magnitude of the treatment effect---from expert opinion in the context of clinical trials with DTEs. We also demonstrated how these elicited distributions can be used to calculate the probability of trial success (assurance \cite{OHagan2005}) for a fixed trial design. In this paper, we extend this work by introducing a novel adaptive trial design framework that directly incorporates elicited prior distributions to more effectively address the challenges associated with DTEs. Specifically, we make two main contributions. First, we extend the assurance framework to adaptive designs with interim analyses, allowing trialists to evaluate how interim looks affect the overall probability of trial success. Second, we develop a Predictive Probability (PP) framework for futility decision-making at interim, in which elicited priors are sequentially updated with trial data to compute the probability that the trial will ultimately succeed, providing a principled and transparent basis for early stopping decisions whilst maintaining conventional frequentist error control for efficacy. We additionally provide a practical calibration procedure for selecting the PP threshold and interim timing, including optimisation with respect to the expected sample size under the null hypothesis.
To support practical implementation, we have developed an open-source R package that allows users to simulate and compare different adaptive trial designs, alongside an R Shiny application based on the widely used \texttt{{rpact}} \cite{Wassmer2024} package. Installation and usage instructions are provided in Appendix A.
In Section \ref{sec:DTE}, we reintroduce the parameterisation and notation for DTEs as outlined in our previous work.\cite{Salsbury2024} In Section \ref{sec:Framework}, we detail our methodology for integrating elicited prior distributions into the planning of interim analyses for trials with DTEs. We present a practical example in Section \ref{sec:example}, illustrating how interim analyses can inform early stopping decisions. We conclude with a discussion in Section \ref{sec:discussion}.

\section{Model Parameterisation and Prior Distributions}\label{sec:DTE}

In this section, we reintroduce the parameterisation and notation for delayed treatment effects (DTEs), as outlined in our previous work.\cite{Salsbury2024} We also describe the prior distributions over the model parameters that are assumed to have been elicited prior to trial planning. 

\subsection{Parameterisation}\label{sec:parameterisation}
Assume we are planning a survival trial with two groups: the control group and the experimental treatment group. We denote the hazard function for the control group as $h_c(t)$ and the hazard function for the experimental treatment group as $h_e(t)$. In a typical survival trial, we assume that patients in the experimental treatment arm may immediately benefit from the intervention compared to those in the control arm; that is, $h_e(t)\leq h_c(t)\forall T$.

In a trial in which a DTE is thought likely to occur, we make a different assumption. We assume that the hazard function for the experimental treatment group is the same as that of the control group until a certain time $T$, which represents the delay in the experimental treatment taking effect. After time $T$, we assume that the experimental treatment group begins to experience some benefit relative to the control group:
\begin{equation}\label{eq:HF}
h_e(t) =
\begin{cases}
h_c(t), & t \leq T \\
h^{*}_e(t), & t > T
\end{cases},
\end{equation}
where $h_e^*(t)\leq h_c(t)$ describes the benefit of the experimental treatment relative to control.\cite{Fine2007}

We suppose that survival times in the control group follow a Weibull distribution with hazard function
\begin{equation}\label{eq:HFc}
h_c(t) = \gamma_c\lambda_c^{\gamma_c} t^{\gamma_c-1}
\end{equation}
and corresponding survival function
\begin{equation}\label{eq:control}
S_c(t) = \exp{-(\lambda_c t)^{\gamma_c}}.
\end{equation}
For the experimental treatment group, we make the assumption that $\gamma_e = \gamma_c$, which implies that proportional hazards hold once the treatment begins to take effect.\cite{Ristl2020} Under this assumption, the post-delay hazard ratio simplifies to a constant, which we denote $\text{HR}^*$:
\begin{equation}\label{eq:HR2}
\text{HR}(t) =
\begin{cases}
1, & t \leq T \\
\left(\frac{\lambda_e}{\lambda_c}\right)^{\gamma_c}, & t > T
\end{cases}
\end{equation}
The survival function for the experimental treatment group is therefore
\begin{equation}\label{eq:treatment}
S_e(t) =
\begin{cases}
\exp{-(\lambda_c t)^{\gamma_c}}, & t \leq T \\
\exp{-(\lambda_c T)^{\gamma_c} - \text{HR}^*\left[(\lambda_c t)^{\gamma_c} - (\lambda_c T)^{\gamma_c}\right]}, & t > T
\end{cases}.
\end{equation}
Under the common shape assumption, the model is fully characterised by four parameters: $\boldsymbol{\theta} = (\lambda_c, \gamma_c, \text{HR}^*, T)$. These are the parameters over which prior distributions are elicited, as described in the following subsections.

\subsection{\texorpdfstring{Prior distributions for $\lambda_c$ and $\gamma_c$}{Prior distributions for lambda\_c, gamma\_c}}\label{sec:controlpriors}
We assume that historical data on the control group intervention are available, from which we derive
\begin{equation*}
\pi(\lambda_c, \gamma_c \mid \boldsymbol{x}_{\text{hist}}),
\end{equation*}
where $\boldsymbol{x}_\text{hist}$ denotes the historical data for the control group. We assume $(\lambda_c, \gamma_c)$ are independent of $(\text{HR}^*, T)$ in the joint prior.

\subsection{\texorpdfstring{Prior distribution for $\mathrm{HR}^{*}$}{Prior distribution for HR*}}

We must account for the possibility that the survival curves may not separate at all, that is, there is no treatment effect of any kind. We define $S$ to be the proposition that the population survival curves separate, and let $P_S$ denote the elicited probability that this proposition is true. The prior distribution for $\text{HR}^*$ is therefore a mixture:
\begin{equation*}
\text{HR}^* \sim
\begin{cases}
1, & \text{with probability } 1 - P_S \\
D_{\text{HR}^*}, & \text{with probability } P_S
\end{cases},
\end{equation*}
where $D_{\text{HR}^*}$ is the elicited distribution of the post-delay hazard ratio, conditional on the survival curves separating.

\subsection{\texorpdfstring{Prior distribution for $T$}{Prior distribution for T}}

The length of the delay $T$ is a parameter that can be intuitively understood and directly elicited from experts. It is only relevant when the survival curves are expected to eventually separate; thus, this parameter is considered conditional on $S$ being true. We additionally incorporate the possibility of there being no delay in the separation of the curves, that is denoted by $P_\text{DTE}$. The prior distribution for $T$ is:
\begin{equation*}
T \mid S \sim
\begin{cases}
0, & \text{with probability } 1 - P_\text{DTE} \\
D_{T}, & \text{with probability } P_\text{DTE}
\end{cases},
\end{equation*}
where $D_{T}$ is the elicited distribution of the length of delay, conditional on a delayed treatment effect occurring.

\subsection{Summary of Prior Distributions}\label{sec:elicitedParameters}
The combination of $P_S$, $P_\text{DTE}$, and the continuous distributions $D_{\text{HR}^*}$ and $D_T$ gives rise to three clinically meaningful scenarios, summarised in Table~\ref{tab:scenarios}.

\begin{table}[h]
\centering
\caption{The three possible scenarios arising from the mixture prior structure.}
\label{tab:scenarios}
\begin{tabular}{llp{12cm}}
\hline
$S$ & DTE & Implication \\
\hline
False & -- & $\text{HR}^*=1$, $T$ irrelevant; no treatment effect of any kind \\
True & False & $\text{HR}^* \sim D_{\text{HR}^*}$, $T=0$; immediate proportional hazards treatment effect \\
True & True & $\text{HR}^* \sim D_{\text{HR}^*}$, $T \sim D_T$; delayed treatment effect \\
\hline
\end{tabular}
\end{table}

In summary, the quantities required to plan a clinical trial in this context are:
\begin{itemize}
\item $\pi(\lambda_c, \gamma_c \mid \boldsymbol{x}_{\text{hist}})$: the joint posterior distribution for the control group Weibull parameters
\item $P_S$: the elicited probability that the survival curves will separate
\item $D_{\text{HR}^*}$: the elicited distribution of the post-delay hazard ratio, conditional on separation
\item $P_\text{DTE}$: the elicited probability of a delayed treatment effect, conditional on separation
\item $D_{T}$: the elicited distribution of the length of delay, conditional on a delayed treatment effect
\end{itemize}

For the remainder of this paper, we assume these quantities have been obtained via the elicitation methods described in Salsbury et al.\cite{Salsbury2024} and use them to plan an adaptive clinical trial.

\section{Assurance with Interim Analyses}\label{sec:Framework}

\subsection{Assurance in the Fixed Design 
Setting}\label{sec:assurance_fixed}

The concept of assurance, introduced by Spiegelhalter et al.\cite{Spiegelhalter1986b}, and coined by O'Hagan et 
al.\cite{OHagan2005}, provides a Bayesian alternative to 
traditional power calculations for clinical trial design. Whereas 
power is computed at a fixed point alternative, assurance accounts 
for uncertainty in the true parameter values by averaging the 
probability of trial success over the prior distribution of the 
model parameters. Formally, the assurance A, which depends on the chosen trial design, is defined as

\begin{equation}\label{eq:assurance}
A = \int_{\boldsymbol{\Theta}} P(\text{reject } H_0 \mid 
\boldsymbol{\theta}) \, \pi(\boldsymbol{\theta}) \, 
d\boldsymbol{\theta},
\end{equation}

\noindent where $\pi(\boldsymbol{\theta})$ is the prior distribution 
over the model parameters $\boldsymbol{\theta} = (\lambda_c, 
\gamma_c, \text{HR}^*, T)$ as described in 
Section~\ref{sec:DTE}, and $P(\text{reject } H_0 \mid 
\boldsymbol{\theta})$ is the probability of rejecting the null 
hypothesis at the final analysis under a fixed trial design, 
conditional on $\boldsymbol{\theta}$. In our previous 
work\cite{Salsbury2024}, we demonstrated how assurance can be 
computed for a fixed trial design in the presence of a delayed 
treatment effect, using the elicited prior distributions described 
in Section~\ref{sec:DTE}. Since a closed-form expression for 
Equation~\ref{eq:assurance} is not available under the piecewise 
Weibull model, assurance is estimated via simulation as

\begin{equation}\label{eq:assurance_est}
\hat{A} = \frac{1}{N} \sum_{i=1}^{N} 
\mathbbm{1}(\text{reject } H_0 \text{ in simulation } i),
\end{equation}

\noindent where each simulated trial is conducted under a draw 
$\boldsymbol{\theta}_i \sim \pi(\boldsymbol{\theta})$.

\subsection{Extension to Adaptive 
Designs}\label{sec:assurance_adaptive}

We now extend the assurance framework to accommodate adaptive trial 
designs with interim analyses. In the adaptive setting, the trial 
may terminate early---either for futility or for efficacy---at one 
of $L$ pre-specified interim looks, or continue to a final 
analysis. We define a trial as \textit{successful} if it results 
in rejection of $H_0$ under any of these scenarios; that is, 
either the trial stops early for efficacy at an interim look, or 
the null hypothesis is rejected at the final analysis having not 
stopped early. Formally, let $\mathcal{W}$ denote the event of a 
successful trial outcome. Then

\begin{equation}
\mathcal{W} =
\left\{
\bigcup_{j=1}^{L}
\{\text{stop for efficacy at look } j\}
\right\}
\cup
\{ \text{reject } H_0 \text{ at final analysis}\},
\end{equation}

\noindent where these events are mutually exclusive by construction; the trial stops at the first look at which a stopping criterion 
is met.

Assurance in the adaptive setting is then defined analogously to 
Equation~\ref{eq:assurance}, with the probability of trial success 
now reflecting the adaptive stopping rules:

\begin{equation}\label{eq:assurance_adaptive}
A = \int_{\boldsymbol{\Theta}} P(\mathcal{W} \mid 
\boldsymbol{\theta}) \, \pi(\boldsymbol{\theta}) \, 
d\boldsymbol{\theta}.
\end{equation}

\noindent Note that Equation~\ref{eq:assurance_adaptive} reduces to 
Equation~\ref{eq:assurance} in the absence of interim analyses, 
confirming that the fixed design assurance is a special case of 
the more general framework presented here.

The probability $P(\mathcal{W} \mid \boldsymbol{\theta})$ depends 
on the stopping rules applied at each interim look. In a group 
sequential design, efficacy stopping at interim look $j$ occurs 
when the test statistic $Z_j$ exceeds the pre-specified efficacy 
boundary $b_j$, which is computed to control the familywise Type I 
error rate at level $\alpha$ via a pre-specified alpha-spending 
function. Futility stopping at interim look $j$ is governed by a 
pre-specified stopping rule $\mathcal{S}$, which is treated as a 
general input to the framework. 

Since a closed-form expression for 
Equation~\ref{eq:assurance_adaptive} is not available under the 
piecewise Weibull model with adaptive stopping rules, assurance is 
estimated via the simulation procedure described in 
Algorithm~\ref{alg:GSDDTE}. Briefly, for each of $N$ iterations, 
a parameter vector $\boldsymbol{\theta}_i$ is drawn from the joint 
prior $\pi(\boldsymbol{\theta})$, a trial is simulated under those 
parameters, the stopping rules are applied at each interim look, 
and the trial outcome $U_i \in \{0, 1\}$ is recorded. The 
assurance estimate $\hat{A}$ is then computed as in 
Equation~\ref{eq:assurance_est}.

A key practical consideration in the adaptive setting concerns the 
timing of interim analyses relative to the delay parameter $T$. If 
an interim look is scheduled before a meaningful proportion of 
patients have passed $T$, the observed data will be largely 
uninformative about $\text{HR}^*$ and $T$, and any futility 
decision based on such data may be unreliable. In practice, 
interim analyses should therefore be scheduled with reference to 
the prior distribution of $T$, to ensure that sufficient 
information about the treatment effect has accrued by the time of 
the look. We discuss this further in the context of the Bayesian 
Predictive Probability stopping rule in 
section~\ref{sec:PP}.

\subsection{Simulation Procedure}\label{sec:simulation}

At a high level, the procedure 
repeatedly draws parameter vectors from the joint prior 
$\pi(\boldsymbol{\theta})$, simulates a complete trial under those 
parameters, including the application of stopping rules at each 
interim look, and records whether the trial was successful. The 
assurance estimate is then the proportion of successful trials 
across all iterations. The full procedure is described in 
Algorithm~\ref{alg:GSDDTE}.

The choice of test statistic $Z(\cdot)$ is left to the 
practitioner. In many Phase III trials, the standard log-rank test 
is used, and the efficacy boundaries $\mathbf{b}$ are computed 
accordingly via a pre-specified alpha-spending function using 
software such as \texttt{rpact}\cite{Wassmer2024}. However, 
alternative test statistics, such as weighted log-rank tests or 
max-combo tests\cite{Mukhopadhyay2020}, may offer power advantages in the 
presence of a delayed treatment effect, and the framework 
accommodates these naturally provided the efficacy boundaries are 
computed to correspond to the chosen statistic. The futility stopping rule $\mathcal{S}$ is similarly treated as a 
general input. Any pre-specified stopping rule may be substituted 
, for example, a standard beta-spending futility boundary. The computational cost of Algorithm~\ref{alg:GSDDTE} depends 
primarily on the number of iterations $N$ and the complexity of 
the futility stopping rule $\mathcal{S}$.

\section{Predictive Probability with Delayed Treatment Effects} \label{sec:PP}

Predictive Probability (PP) provides a model-based framework for interim decision-making, but its application requires explicit specification of the data-generating process for both the observed and future outcomes. In the presence of delayed treatment effects, this step becomes non-standard: the likelihood must accommodate a piecewise treatment hazard, and the predictive simulations must account for uncertainty in both the magnitude and timing of the delayed effect. This section develops the extension of PP to delayed-effect survival models.

At an interim analysis, the elicited prior distributions are updated using the observed data to obtain joint posterior distributions. Posterior draws are then propagated forward to simulate the unobserved follow-up, reconstruct the final analysis, and evaluate the prespecified success criterion. Averaging these indicators yields the PP: the probability of ultimately achieving the primary endpoint conditional on the interim information and the prior distributions.

To implement this procedure, we formulate explicit likelihoods for the delayed-effect models. As introduced in Section \ref{sec:parameterisation}, we consider Weibull control arms with a piecewise treatment hazard to represent the delay. This likelihood, combined with the elicited priors, permit posterior computation via MCMC and form the basis of the predictive probability calculations.

\subsection{Likelihood}\label{sec:likelihood}

Let $x_i$ denote the observed follow-up time, $y_i \in \{0,1\}$ the event 
indicator, and $z_i \in \{0,1\}$ the treatment indicator. Under a delayed 
treatment effect with delay duration $T$, treated subjects contribute either 
through the pre-delay or post-delay hazard depending on whether $x_i \le T$ 
or $x_i > T$. For censored observations ($y_i = 0$), only the cumulative hazard 
contributes.

Define the index sets

\begin{itemize}
    \item $\mathcal{C} = \{ i : z_i = 0 \}$ \\
    \item $\mathcal{T}_{\le T} = \{ i : z_i = 1,\, x_i \le T \}$ \\
    \item $\mathcal{T}_{>T} = \{ i : z_i = 1,\, x_i > T \}$
    
\end{itemize}

For all $i \in \mathcal{C} \cup \mathcal{T}_{\le T}$, the likelihood contribution is
\[
L_i
= \left( \gamma_c \lambda_c^{\gamma_c} x_i^{\gamma_c - 1} \right)^{y_i}
  \exp\!\left( -(\lambda_c x_i)^{\gamma_c} \right).
\]

For $i \in \mathcal{T}_{> T}$, the cumulative hazard decomposes into a 
pre-delay segment plus a post-delay segment scaled by $\text{HR}^*$:

\begin{align*}
    H_i(x_i)  &= H_C(T) + \text{HR}^*\bigl[ H_C(x_i) - H_C(T) \bigr] \\
    &= (\lambda_c T)^{\gamma_c}
  + \text{HR}^*\bigl( (\lambda_c x_i)^{\gamma_c}
                      - (\lambda_c T)^{\gamma_c} \bigr)
\end{align*}

The likelihood contribution is therefore
\[
\begin{aligned}
L_i
&= \left( \text{HR}^* \gamma_c \lambda_c^{\gamma_c} x_i^{\gamma_c - 1} \right)^{y_i} \\
&\quad \times \exp\!\Big[
-(\lambda_c T)^{\gamma_c}
-\text{HR}^* \bigl(
(\lambda_c x_i)^{\gamma_c}
- (\lambda_c T)^{\gamma_c}
\bigr)
\Big].
\end{aligned}
\]

Let $\boldsymbol{\theta} = (\lambda_c, \gamma_c, \text{HR}^*, T)$ denote the
parameter vector. The full likelihood is
\[
\begin{aligned}
\mathcal{L}(\boldsymbol{\theta} \mid \mathcal{D})
&=
\prod_{i \in \mathcal{C} \cup \mathcal{T}_{\le T}}
\Bigg[
\left( \gamma_c \lambda_c^{\gamma_c} x_i^{\gamma_c-1} \right)^{y_i} \\
&\qquad\qquad\qquad \times
\exp\!\Big( -(\lambda_c x_i)^{\gamma_c} \Big)
\Bigg] \\
&\quad \times
\prod_{i \in \mathcal{T}_{> T}}
\Bigg[
\left( \text{HR}^* \gamma_c \lambda_c^{\gamma_c} x_i^{\gamma_c-1} \right)^{y_i} \\
&\qquad\qquad\qquad \times
\exp\!\Big(
-(\lambda_c T)^{\gamma_c} \\
&\qquad\qquad\qquad\qquad
-\text{HR}^* \bigl(
(\lambda_c x_i)^{\gamma_c}
- (\lambda_c T)^{\gamma_c}
\bigr)
\Big)
\Bigg].
\end{aligned}
\]

\subsection{Posterior Updating Under Delayed Treatment Effects}
\label{sec:posterior_dte}

Given the likelihood expressions defined above, Bayesian updating proceeds by 
combining the likelihood with prior distributions on all model parameters. At the interim 
analysis, the posterior is
\[
p(\boldsymbol{\theta} \mid \mathcal{D}_{\mathrm{int}})
\;\propto\;
\mathcal{L}(\boldsymbol{\theta} \mid \mathcal{D}_{\mathrm{int}})
\, p(\boldsymbol{\theta}),
\]
where $\mathcal{D}_{\mathrm{int}}$ denotes the event and censoring times 
observed up to the interim look.

Posterior inference is obtained by sampling from 
$p(\boldsymbol{\theta} \mid \mathcal{D}_{\mathrm{int}})$ using 
\texttt{rjags}\citep{rjags}.

Convergence is assessed using standard diagnostics; trace plots, effective 
sample sizes, and the $\widehat{R}$ statistic. The resulting posterior sample,
\[
\{\boldsymbol{\theta}^{(m)} : m = 1,\ldots,M\},
\]
provides the basis for the predictive probability calculations 
described in the next section.

\subsection{Calculating PP with Delayed Treatment Effects}

Following posterior updating at the interim analysis, the goal is to quantify
the probability that the final analysis will reject $H_0$, accounting for
uncertainty in both the model parameters and all future unobserved outcomes.
Let $\boldsymbol{\theta}$ denote the full parameter vector and
$\tilde{\mathcal{D}}$ the unobserved survival and censoring outcomes between
the interim and final analyses.

Unlike standard proportional-hazards settings, computing PP under delayed
treatment effects is non-trivial. At interim, individuals belong to distinct
risk-set categories, each governed by different hazard functions:
\begin{itemize}
    \item patients not yet enrolled (both arms),
    \item censored control patients,
    \item censored treated patients with $x_i < T$ (pre-delay region),
    \item censored treated patients with $x_i > T$ (post-delay region),
    \item patients who have already experienced the event.
\end{itemize}
Future event times must therefore be simulated under a piecewise hazard
structure that depends jointly on (i) treatment assignment, (ii) accumulated
follow-up relative to the delay time~$T$, and (iii) the sampled parameter
vector $\boldsymbol{\theta}^{(m)}$. This heterogeneity in data-generating 
mechanisms is what makes predictive calculations under delayed effects
substantially more complex than in models with proportional hazards.

For the reconstructed full dataset
\[
\mathcal{D}_{\mathrm{int}} \cup \tilde{\mathcal{D}}^{(m)},
\]
the final analysis is performed exactly as specified in the statistical 
analysis plan. Define
\[
\mathbb{I}^{(m)}
=
\mathbb{I}\!\left\{
  \text{Final analysis rejects } H_0
  \;\middle|\;
  \mathcal{D}_{\mathrm{int}},
  \tilde{\mathcal{D}}^{(m)},
  \boldsymbol{\theta}^{(m)}
\right\}.
\]

The predictive probability is estimated by
\[
\widehat{\mathrm{PP}}
=
\frac{1}{M} \sum_{m=1}^M \mathbb{I}^{(m)},
\]
based on posterior draws 
$\{\boldsymbol{\theta}^{(m)} : m=1,\ldots,M\}$.

A simulation-based algorithm for computing PP under delayed effects is:

\begin{enumerate}
  \item Draw $\boldsymbol{\theta}^{(m)} \sim 
        p(\boldsymbol{\theta} \mid \mathcal{D}_{\mathrm{int}})$ using MCMC.
  \item For each draw, simulate future survival and censoring times from 
        $p(\tilde{\mathcal{D}} \mid \boldsymbol{\theta}^{(m)})$, applying the 
        correct treatment-specific and delay-dependent hazard to each risk-set 
        category.
  \item Form the completed dataset 
        $\mathcal{D}_{\mathrm{int}} \cup \tilde{\mathcal{D}}^{(m)}$ and compute 
        the final-analysis test statistic.
  \item Record $\mathbb{I}^{(m)}$, indicating whether the final analysis rejects 
        $H_0$.
  \item Average the indicators to obtain $\widehat{\mathrm{PP}}$.
\end{enumerate}

This Monte Carlo estimator converges almost surely to the true predictive 
probability as $M \to \infty$. In practice, 
$M$ in the range $1{,}000$--$10{,}000$ yields adequate numerical stability for 
interim decision-making.

\subsection{Futility Stopping 
Rule}\label{sec:futility}

Having obtained the predictive probability estimate 
$\widehat{\text{PP}}$ at the interim analysis, we now define the 
futility stopping rule $\mathcal{S}$ introduced in 
Algorithm~\ref{alg:GSDDTE}. We stop for futility at the interim 
look if the predictive probability of trial success falls below a 
pre-specified threshold $\lambda \in (0, 1)$:

\begin{equation}\label{eq:futility}
\mathcal{S}: \quad \text{stop for futility if } 
\widehat{\text{PP}} < \lambda.
\end{equation}

\noindent Intuitively, if the probability of ultimately rejecting 
$H_0$ at the final analysis,  given the interim data and our 
updated beliefs about the model parameters is sufficiently low, 
there is little justification for continuing the trial. Early 
stopping for futility in this case reduces patient burden and 
resource expenditure without compromising the scientific 
integrity of the trial.

It is important to emphasise that this framework is proposed 
specifically for \textit{futility} stopping in a confirmatory 
Phase III setting. Efficacy stopping remains governed by the 
pre-specified GSD efficacy boundaries $\mathbf{b}$, computed via 
a standard alpha-spending function to control the familywise Type 
I error rate at level $\alpha$. The PP-based futility rule does 
not affect Type I error control--- stopping early for futility is 
conservative from a regulatory perspective, as it reduces the 
probability of a false positive result. Nevertheless, the impact 
of the futility rule on the overall operating characteristics of 
the trial, including Type I error and power, should be 
evaluated explicitly via the calibration procedure described in 
Section~\ref{sec:calibration}.

The threshold $\lambda$ is a key design parameter that governs 
the aggressiveness of the futility rule. A higher value of 
$\lambda$ leads to earlier and more frequent futility stopping, 
reducing expected sample size under the null but potentially 
increasing the risk of stopping a genuinely effective trial 
prematurely. Conversely, a lower value of $\lambda$ is more 
conservative, allowing more trials to continue to the final 
analysis. The choice of $\lambda$ should therefore be made with 
reference to the trial's operating characteristics under a range 
of clinically meaningful scenarios, as described in 
Section~\ref{sec:calibration}.

\subsection{Calibration of the PP 
Threshold}\label{sec:calibration}

The futility stopping rule defined in 
Section~\ref{sec:futility} requires the specification of two 
design parameters: the interim timing and the threshold $\lambda$. 
Both should be determined prior to the trial commencing, as part 
of the trial design process, and pre-specified in the statistical 
analysis plan. We propose a two-step calibration procedure for 
selecting these parameters, which we describe below.

\subsubsection{Step 1: Selecting the Interim 
Timing}\label{sec:timing}

The interim timing should be chosen to ensure that the PP is 
sufficiently informative to support a meaningful futility 
decision. If the interim occurs too early, i.e. before a meaningful 
proportion of patients have passed the delay time $T$, the 
observed data will be largely uninformative about $\text{HR}^*$ 
and $T$, and the resulting PP values will be concentrated around 
intermediate values, reflecting prior uncertainty rather than 
accumulated evidence. We propose selecting the interim timing by 
examining the distribution of $\widehat{\text{PP}}$ under the 
prior predictive distribution at a range of candidate timings, 
as follows:

\begin{enumerate}
    \item Specify a grid of candidate interim timings 
    $t_1 < t_2 < \cdots < t_K$, expressed as information 
    fractions of the maximum event count $E$.
    
    \item For each candidate timing $t_k$:
    \begin{enumerate}[label=(\roman*)]
        \item Draw $N_{\text{cal}}$ parameter vectors 
        $\boldsymbol{\theta}^{(n)} \sim \pi(\boldsymbol{\theta})$ 
        from the joint prior.
        \item For each draw, simulate a trial dataset up to 
        timing $t_k$ using Algorithm~\ref{alg:GSDDTE}.
        \item Compute $\widehat{\text{PP}}^{(n)}$ for each 
        simulated dataset using the procedure described in 
        Section~\ref{sec:PP}.
        \item Plot the distribution of 
        $\{\widehat{\text{PP}}^{(n)}\}_{n=1}^{N_{\text{cal}}}$ 
        as a histogram.
    \end{enumerate}
    
    \item Select the earliest timing $t^*$ at which the 
    histogram of $\widehat{\text{PP}}$ values is sufficiently 
    polarised towards 0 and 1, indicating that the interim data 
    are informative enough to distinguish promising from 
    unpromising trials.
\end{enumerate}

\noindent In practice, $N_{\text{cal}} = 500$ datasets is 
sufficient to produce stable histograms. We illustrate this 
procedure in the case study of Section~\ref{sec:example}.

\subsubsection{Step 2: Selecting the Threshold}\label{sec:threshold}

Having fixed the interim timing at $t^*$, the threshold $\lambda$ 
is selected by evaluating the operating characteristics of the 
trial design under a range of clinically meaningful scenarios. 
Each scenario is defined by a fixed parameter vector 
$\boldsymbol{\theta}_s = (\lambda_{c,s}, \gamma_{c,s}, 
\text{HR}^*_s, T_s)$, chosen to reflect a specific clinical 
hypothesis of interest. We recommend including at minimum:

\begin{itemize}
    \item \textit{Null scenario} ($\text{HR}^* = 1$): verifies 
    that the futility rule does not inflate the Type I error rate. 
    Note that under the null, $\text{HR}^* = 1$ and the survival 
    distributions are identical on both arms regardless of $T$; 
    the delay parameter is therefore irrelevant under $H_0$.
    
    \item \textit{Proportional hazards alternative} ($T = 0$, 
    $\text{HR}^* < 1$): a useful benchmark reflecting the 
    standard proportional hazards assumption.
    
    \item \textit{Delayed treatment effect alternatives}: one or 
    more scenarios with $T > 0$ and $\text{HR}^* < 1$, reflecting 
    clinically plausible delay lengths. The specific values of $T$ 
    and $\text{HR}^*$ should be chosen with reference to the 
    elicited prior distributions. For example, the prior median 
    of each parameter provides a natural and principled choice.
\end{itemize}

\noindent For each scenario $s$ and each candidate threshold 
$\lambda$, the following operating characteristics are estimated 
via simulation of $N_{\text{cal}}$ trials under 
$\boldsymbol{\theta}_s$, applying the PP futility rule with 
threshold $\lambda$ at interim timing $t^*$:

\begin{itemize}
    \item \textit{Probability of rejecting $H_0$}: the proportion 
    of simulated trials that reject $H_0$ at the final analysis. 
    Under the null scenario this quantity is the Type I error; 
    under each alternative scenario it is the power.
    
    \item \textit{Expected sample size} (ESS): the expected number 
    of patients enrolled at the point of trial termination, 
    averaged across simulations. This reflects the efficiency of 
    the design under each scenario, trials stopping early for 
    futility will have a lower ESS than those continuing to the 
    final analysis.
\end{itemize}

\noindent The threshold $\lambda$ is then selected by examining these operating characteristics across all scenarios and choosing a value that achieves acceptable Type I error control and satisfactory power under the alternative scenarios of interest. This is necessarily a judgement-based decision that reflects the trialist's priorities --- a higher $\lambda$ increases efficiency under the null at the cost of greater risk of false futility stopping under the alternatives, and vice versa. More formally, the threshold selection can be framed as a utility maximisation problem, in which a utility function explicitly trades off Type I error, power, and expected sample size across scenarios; the specification of such a utility is inherently context-dependent and is left to the trialist.

As a practically important special case, consider minimising the expected sample size under the null, subject to power remaining above some acceptable level $1 - \beta^*$ under each alternative scenario of interest. This can be solved via the following procedure:

\begin{enumerate}
    \item Specify a grid of candidate thresholds $\lambda_1 < \lambda_2 < \cdots < \lambda_K$ and a minimum acceptable power level $1 - \beta^*$.
    \item For each $\lambda_k$, simulate $N_{\text{cal}}$ trials under the null scenario and each alternative scenario $s \in \mathcal{S}_{\text{alt}}$, and estimate $\mathrm{ESS}(\lambda_k \mid H_0)$ and $\mathrm{Power}(\lambda_k \mid \boldsymbol{\theta}_s)$.
    \item Discard any $\lambda_k$ for which $\mathrm{Power}(\lambda_k \mid \boldsymbol{\theta}_s) < 1 - \beta^*$ for any $s \in \mathcal{S}_{\text{alt}}$.
    \item Select
    \begin{equation}
        \lambda^* = \max\bigl\{\lambda_k : \mathrm{Power}(\lambda_k \mid \boldsymbol{\theta}_s) \geq 1 - \beta^* \; \forall s \in \mathcal{S}_{\text{alt}}\bigr\},
    \end{equation}
    i.e.\ the largest feasible threshold, which minimises $\mathrm{ESS}$ under the null since more aggressive futility stopping corresponds to larger $\lambda$.
\end{enumerate}

We illustrate the judgement-based approach in the case study of Section~\ref{sec:example}. The selected design parameters $(t^*, \lambda^*)$ should be pre-specified in the statistical analysis plan prior to the commencement of the trial.

It is important to distinguish the calibration procedure from 
the assurance calculation of Section~\ref{sec:Framework}. 
Assurance averages over the full joint prior 
$\pi(\boldsymbol{\theta})$, providing a Bayesian summary of the 
trial's overall probability of success under parameter 
uncertainty. Calibration, by contrast, evaluates operating 
characteristics at fixed point scenarios, providing a 
frequentist assessment of the design's behaviour under specific 
hypotheses. Rather than viewing these as competing approaches, 
we recommend reporting both, as together they provide a richer and 
more complete picture of the trial design than either perspective 
alone.

\subsection{Sensitivity Analyses}\label{sec:sens_analyses}

Where prior distributions are based on expert judgement, it is desirable to understand the sensitivity of any conclusions or decisions to such judgements. One could simply re-run the entire analysis with a different prior distribution, but we do not demonstrate this in our example in Section \ref{sec:example}.  We instead discuss (and illustrate) how sensitivities to prior judgements can be inferred from a full simulation with a single choice of prior distribution, through categorising and reporting particular simulation outputs.

Note that the prior distribution is used in two places in our framework: in the data-generating mechanism (to calculate assurance/PoS), and
at the interim stage (where they are updated with the observed data to yield the posterior probability). We consider sensitivity analyses for each of these in turn.

\subsubsection{Sensitivity to the data-generating mechanism}

Several sensitivity checks can be performed on the assurance without rerunning the 
simulation, by retaining intermediate quantities from the original run and reweighting 
post hoc.

\paragraph{Sensitivity to $P_S$ and $P_{\text{DTE}}$.}
Recall from Table~\ref{tab:scenarios} that $P_S$ and $P_{\text{DTE}}$ define three
underlying scenarios. Using stratified sampling, a fixed proportion of samples is drawn
from each scenario, so that the scenario-specific assurances $\mathcal{A}_{\text{null}}$,
$\mathcal{A}_{\text{delay}}$, and $\mathcal{A}_{\text{PH}}$ are estimated independently.
The overall assurance is then:
\begin{equation}\label{eq:weighted_ass}
    \mathcal{A}_{\text{overall}} = (1 - P_S)\,\mathcal{A}_{\text{null}}
    + P_S P_{\text{DTE}}\,\mathcal{A}_{\text{delay}}
    + P_S(1 - P_{\text{DTE}})\,\mathcal{A}_{\text{PH}}.
\end{equation}
Since the scenario-specific assurances are estimated independently, sensitivity to $P_S$
or $P_{\text{DTE}}$ can be assessed by substituting revised values directly into
Equation~\eqref{eq:weighted_ass}, without rerunning the simulation. This approach extends
naturally to other operating characteristics of interest---such as trial duration, sample
size, or probability of stopping for futility---by substituting the relevant quantity
for $\mathcal{A}$ throughout.

\paragraph{Parameter distributions $\mathcal{D}_{\text{HR}^*}$ and $\mathcal{D}_{T}$.}
A similar post hoc sensitivity analysis is available for the parameter distributions. 
Here, we recommend retaining the sampled values of $\text{HR}^*$ and $T$, and 
stratifying by quantile of the respective marginal distribution. Let $\mathcal{A}_{Q_k}$ 
denote the assurance contribution from samples falling in the $k$-th quartile of 
$\mathcal{D}_{\text{HR}^*}$, so that:

\begin{equation}\label{eq:weighted_ass2}
    \mathcal{A}_{\text{overall}} = \sum_{k=1}^{4} \mathcal{A}_{Q_k},
\end{equation}

where each $\mathcal{A}_{Q_k}$ implicitly carries weight $\frac{1}{4}$, since 
each quartile contains an equal proportion of the samples by construction. 
Stratification by quartile is a natural choice, reflecting how experts typically 
reason during elicitation, though the approach generalises to any partition of the 
distribution. To assess sensitivity to over- or under-confidence in elicitation, one 
can reweight the quartile contributions: downweighting the tails corresponds to a more 
concentrated (overconfident) distribution, whilst upweighting them corresponds to a 
more diffuse one. Formally, one replaces the uniform quartile weights $\frac{1}{4}$ 
with alternative weights $\tilde{\omega}_k \geq 0$, $\sum_k \tilde{\omega}_k = 1$, 
giving:

\begin{equation}\label{eq:weighted_ass3}
    \tilde{\mathcal{A}}_{\text{overall}} = \sum_{k=1}^{4} \tilde{\omega}_k \cdot 
    4\,\mathcal{A}_{Q_k}.
\end{equation}

We note that the analyses for $P_S$/$P_{\text{DTE}}$ and for 
$\mathcal{D}_{\text{HR}^*}$/$\mathcal{D}_T$ can in principle be applied simultaneously. 
Since the scenario indicators ($S$, $\text{DTE}$) and the continuous parameters 
($\text{HR}^*$, $T$) are sampled independently in the data-generating mechanism, the 
two reweightings are multiplicatively separable: the combined sensitivity weight for 
sample $i$ is simply $\omega_{s(i)} \cdot \tilde{\omega}_{k(i)}$, where $s(i)$ denotes 
the scenario and $k(i)$ the quartile stratum of that sample.

Finally, the quartile sensitivity described above operates on the marginal distributions 
of $\text{HR}^*$ and $T$ separately. If the two parameters have a non-trivial dependence 
structure, a joint sensitivity analysis would in principle be more appropriate, though 
this substantially increases complexity. In practice, marginal analyses provide a 
reasonable approximation provided the dependence is weak, and we recommend reporting 
them alongside the assumed correlation structure so that readers can judge their 
adequacy.

\subsubsection{Sensitivity to the interim prior}

At the interim, the prior $\pi(\theta)$ is updated with the observed data 
$\mathbf{x}_{\text{int}}$ to yield the posterior $\pi(\theta \mid \mathbf{x}_{\text{int}})$, 
from which the posterior probability PP is computed. A sensitivity analysis here asks: 
how would PP change if a different prior $\tilde{\pi}(\theta)$ had been specified?

Again, we propose an approach that avoids rerunning the simulation, by exploiting 
importance reweighting. Suppose one retains the posterior samples 
$\theta^{(1)}, \ldots, \theta^{(N)}$ from the original interim update. The posterior 
under an alternative prior $\tilde{\pi}(\theta)$ can be approximated without rerunning 
the update, by reweighting the existing samples:

\begin{equation}\label{eq:IS_weights}
    w^{(i)} = \frac{\tilde{\pi}(\theta^{(i)})}{\pi(\theta^{(i)})}, \qquad 
    \bar{w}^{(i)} = \frac{w^{(i)}}{\sum_{j=1}^{N} w^{(j)}},
\end{equation}

so that the reweighted samples $\{\theta^{(i)}, \bar{w}^{(i)}\}$ approximate 
$\tilde{\pi}(\theta \mid \mathbf{x}_{\text{int}})$. The PP under the alternative prior 
is then a reweighted average of the original PP contributions, requiring no additional 
simulation. This is directly analogous to the reweighting approach used for the 
data-generating mechanism above.

A natural one-dimensional sensitivity sweep is obtained by embedding the prior in a 
power prior framework~\citep{Ibrahim2000}, replacing $\pi(\theta)$ with 
$\pi(\theta)^{\alpha_0}$ for $\alpha_0 \in [0, 1]$. Here $\alpha_0 = 1$ recovers the 
original prior, whilst $\alpha_0 \to 0$ yields a flat prior (data only). The importance 
weights in this case are:

\begin{equation}\label{eq:power_prior_weights}
    w^{(i)} \propto \pi(\theta^{(i)})^{\tilde{\alpha}_0 - 1},
\end{equation}

where $\tilde{\alpha}_0$ is the alternative power. Sweeping $\tilde{\alpha}_0$ over 
$[0,1]$ and plotting the resulting PP provides an intuitive summary of how sensitive 
the interim decision is to the strength of the prior.

It should be noted that importance reweighting degrades when $\tilde{\pi}$ and $\pi$ 
differ substantially, as the importance weights become highly variable. A standard 
diagnostic is the effective sample size:

\begin{equation}\label{eq:ESS}
    N_{\text{eff}} = \frac{\left(\sum_{i=1}^{N} w^{(i)}\right)^2}{\sum_{i=1}^{N} 
    \left(w^{(i)}\right)^2},
\end{equation}

which approaches $N$ when $\tilde{\pi} \approx \pi$ and degrades toward 1 when the 
weights are highly concentrated. We recommend reporting $N_{\text{eff}}$ alongside any 
importance-reweighted sensitivity result, and treating conclusions with caution when 
$N_{\text{eff}} / N$ is small.

\section{Example}\label{sec:example}

We illustrate the proposed framework using a hypothetical 
example. We consider a 
two-arm Phase III superiority trial testing a new 
immunotherapy versus docetaxel in patients with advanced 
non-small-cell lung cancer (NSCLC). The primary endpoint is 
overall survival, we assume uniform recruitment over 24 months and 1:1 allocation.

\subsection{Prior Distributions}

The control arm parameters $(\lambda_c, \gamma_c)$ are 
assigned non-informative priors updated using pooled 
reconstructed individual patient data from three published 
docetaxel trials: ZODIAC\cite{Herbst2010}, 
REVEL\cite{Garon2014}, and INTEREST\cite{Kim2008}, as 
described in Salsbury et al.\cite{Salsbury2024} For this 
paper, the pooled data are fitted under a Weibull model, 
yielding a joint posterior $\pi(\lambda_c, \gamma_c \mid 
\boldsymbol{x}_{\text{hist}})$ used as the prior for the 
control arm parameters. 

The remaining prior distributions are assumed to have been elicited from a hypothetical expert. The expert is assumed to have specified $P_S = 0.9$ and $P_{\text{DTE}} = 0.7$. Conditional on a delay occurring, the expert specifies the 25th, 50th, and 75th percentiles of the delay time as 2, 3, and 4 months respectively, which implies $T \sim \text{Gamma}(4.09, 1.28)$. For the post-delay hazard ratio, the expert specifies the 25th, 50th, and 75th percentiles as 0.70, 0.80, and 0.85 respectively, implying $\text{HR}^* \sim \text{Gamma}(45.19, 57.20)$.

These priors induce uncertainty not only about the magnitude of the treatment effect but also about its timing, and therefore play a central role in shaping the posterior distribution at interim and the predictive probability of ultimate success. In the following subsections, we show how these elicited components are integrated into (i) posterior updating at the interim analysis, (ii) simulation-based calibration of the PP futility threshold, and (iii) full forward simulation to evaluate the design's operating characteristics.

\subsection{No Interim Analysis}

We begin by examining a fixed design with no interim monitoring, 
which serves as a baseline against which the impact of introducing 
adaptive decision rules can be assessed. The control arm parameters 
are fixed at the posterior means $\lambda_c = 0.0745$ and 
$\gamma_c = 1.211$, derived from the joint posterior 
$\pi(\lambda_c, \gamma_c \mid \boldsymbol{x}_{\text{hist}})$, 
corresponding to a median control arm survival of approximately 
10 months. Under this parameterisation, approximately 840 events 
are required to detect a hazard ratio of 0.80 with 90\% power 
at a one-sided significance level of $\alpha = 0.025$, assuming 
proportional hazards. We enrol 600 patients per arm over a 
24-month uniform recruitment period, giving a total sample size 
of 1{,}200 patients. Since no interim analyses are performed, 
the sample size remains fixed throughout and the mean trial 
duration is approximately 31 months. These operating 
characteristics provide a reference point for interpreting the 
adaptive designs that follow, allowing direct comparison of how 
interim futility monitoring affects power, expected sample size, 
and trial duration.

\subsection{Adaptive Design Using Predictive 
Probability}\label{sec:adaptive_example}

Having established the baseline fixed design, we now consider an 
adaptive alternative in which a single interim futility analysis 
is based on the predictive probability (PP) of eventual success. 
Control of the one-sided Type~I error rate at $2.5\%$ is 
maintained through a group sequential efficacy boundary, yielding 
a hybrid Bayesian--frequentist monitoring strategy. Construction 
of this adaptive design follows the three-step workflow described 
in Section~\ref{sec:calibration}:

\begin{enumerate}
    \item \textbf{Selecting the timing of the PP interim 
    analysis}, ensuring sufficient information to produce a 
    stable and informative predictive probability while retaining 
    the potential for meaningful efficiency gains;
    \item \textbf{Calibrating the PP-based futility threshold 
    $\lambda$}, balancing early stopping for unpromising scenarios 
    against preservation of power when treatment effects are 
    plausible; and
    \item \textbf{Evaluating the complete design}, embedding both 
    the PP futility rule and group sequential efficacy monitoring 
    across a range of clinically relevant data-generating 
    scenarios.
\end{enumerate}

\subsubsection{Step 1: Choosing the Timing of the PP 
Look}\label{sec:example_timing}

Candidate information fractions between 0.20 and 0.80 (in 
increments of 0.10) were evaluated following the procedure 
described in Section~\ref{sec:timing}. For each candidate timing, 
500 interim datasets were generated from the prior predictive 
distribution under the delayed treatment effect model, the 
posterior was updated via MCMC, and the predictive probability 
(PP) of ultimate trial success was computed.

\begin{figure}[htbp]
\centering
\includegraphics[width=0.7\textwidth]{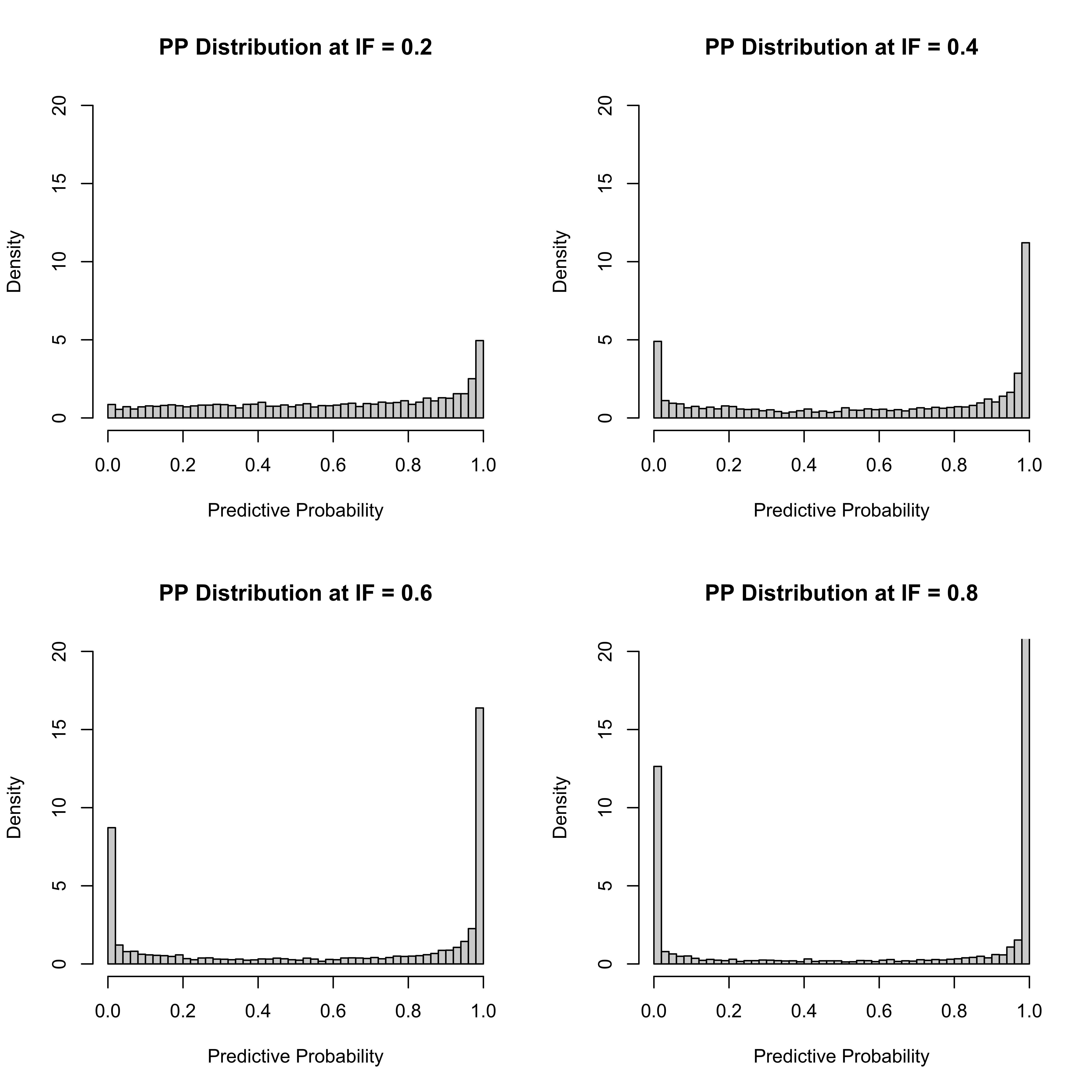}
\caption{Histograms of the predictive probability (PP) at 
selected information fractions. Early looks yield PP 
distributions concentrated near 0.5, reflecting limited 
information, whereas mid-trial looks produce more polarised 
distributions with greater mass near 0 and 1, indicating 
improved discriminatory ability.}
\label{fig:PP_timing_hists}
\end{figure}

Figure~\ref{fig:PP_timing_hists} displays histograms of the PP 
values at selected information fractions. At early information 
fractions, the PP distributions are approximately uniform, 
reflecting the limited information available and the consequent 
dominance of the prior; the interim data are insufficient to 
discriminate reliably between trials that will ultimately 
succeed and those that will not. As the information fraction 
increases, the distributions become increasingly bimodal, with 
mass concentrating near 0 and 1. This reflects the growing 
ability of the accumulated data to distinguish between 
scenarios: trials proceeding under a meaningful treatment 
effect yield PP values close to 1, whilst those under the null 
or a weak effect yield values close to 0. This behaviour is 
consistent with the monotone increase in $\mathcal{I}(t)$ 
reported in Table~\ref{tab:PP_informativeness_timings}.

To quantify the informativeness of each candidate timing, we 
computed
\begin{equation*}
\mathcal{I}(t) = P\!\left(\widehat{\text{PP}} < 0.10 \ 
\text{or} \ \widehat{\text{PP}} > 0.90\right),
\end{equation*}
the probability that the interim analysis yields a near-certain 
futility or success decision. The thresholds 0.10 and 0.90 are 
chosen for illustrative purposes; in practice these could be 
adjusted to reflect the trialist's preferences. 
Table~\ref{tab:PP_informativeness_timings} summarises 
$\mathcal{I}(t)$ and the corresponding expected calendar times 
across all candidate information fractions.

\begin{table}[ht]
\centering
\caption{Informativeness $\mathcal{I}(t)$ of the predictive probability at candidate information fractions, together with the corresponding expected calendar time and sample size at the interim look.}
\label{tab:PP_informativeness_timings}
\begin{tabular}{lccccccc}
\toprule
\textbf{IF} & 0.20 & 0.30 & 0.40 & 0.50 & 0.60 & 0.70 & 0.80 \\
\midrule
$\mathcal{I}(t)$ 
& 0.29 & 0.44 & 0.52 & 0.60 & 0.67 & 0.73 & 0.79 \\
Time (months) 
& 11.8 & 14.7 & 17.2 & 19.5 & 21.6 & 23.7 & 25.8 \\
Sample Size 
& 592.1 & 734.5 & 859.4 & 974.7 & 1082.6 & 1177.2 & 1199.9 \\
\bottomrule
\end{tabular}
\end{table}

Table~\ref{tab:PP_informativeness_timings} shows that 
$\mathcal{I}(t)$ increases monotonically with information 
fraction, as expected. However, the gains become more modest 
at higher information fractions: moving from IF $= 0.50$ to 
IF $= 0.80$ increases $\mathcal{I}(t)$ from $0.60$ to $0.79$, 
whilst adding approximately 6 months and over 200 additional 
patients to the expected time and sample size at the interim. 
The sample size required at the interim also begins to plateau 
beyond IF $= 0.70$ (1177 vs.\ 1199 patients at IF $= 0.80$), 
suggesting limited additional information is being gained in 
that range.

An important operational consideration is that recruitment is 
expected to complete at approximately 24 months. For a futility 
stop to yield meaningful operational gains---for example, by 
halting recruitment and reducing the number of patients exposed 
to an ineffective treatment---the interim analysis must occur 
whilst recruitment is still ongoing. An interim at IF $= 0.50$ 
corresponds to an expected calendar time of 19.5 months, 
comfortably within the recruitment window, whereas later looks 
increasingly risk occurring after recruitment has closed, 
diminishing their practical value.

Taken together, an information fraction of $0.50$ was selected 
as offering a reasonable balance between the informativeness of 
the PP look, the precision of the futility decision, and the 
operational utility of an early stop. This corresponds to an 
expected interim at 19.5 months, with a PP informativeness of 
$\mathcal{I}(0.50) = 0.60$.

\subsubsection{Step 2: Calibration of the PP Futility Threshold}

With the interim timing fixed at $\mathrm{IF} = 0.50$, we 
calibrated the futility threshold $\lambda$ following the procedure 
in Section~\ref{sec:threshold}. To characterise the behaviour of 
the PP under different treatment-effect assumptions, 2{,}000 interim 
datasets were simulated under three fixed-effect scenarios, using 
control parameters $\lambda_c = 0.0745$ and $\gamma_c = 1.211$:
\begin{enumerate}
    \item \textbf{Null}: no treatment effect ($\mathrm{HR} = 1$).
    \item \textbf{Delayed effect}: 3-month delay, post-delay 
    $\mathrm{HR} = 0.8$.
    \item \textbf{Immediate effect}: proportional hazards with 
    $\mathrm{HR} = 0.8$.
\end{enumerate}

\begin{figure}[htbp]
\centering
\includegraphics[width=0.7\textwidth]{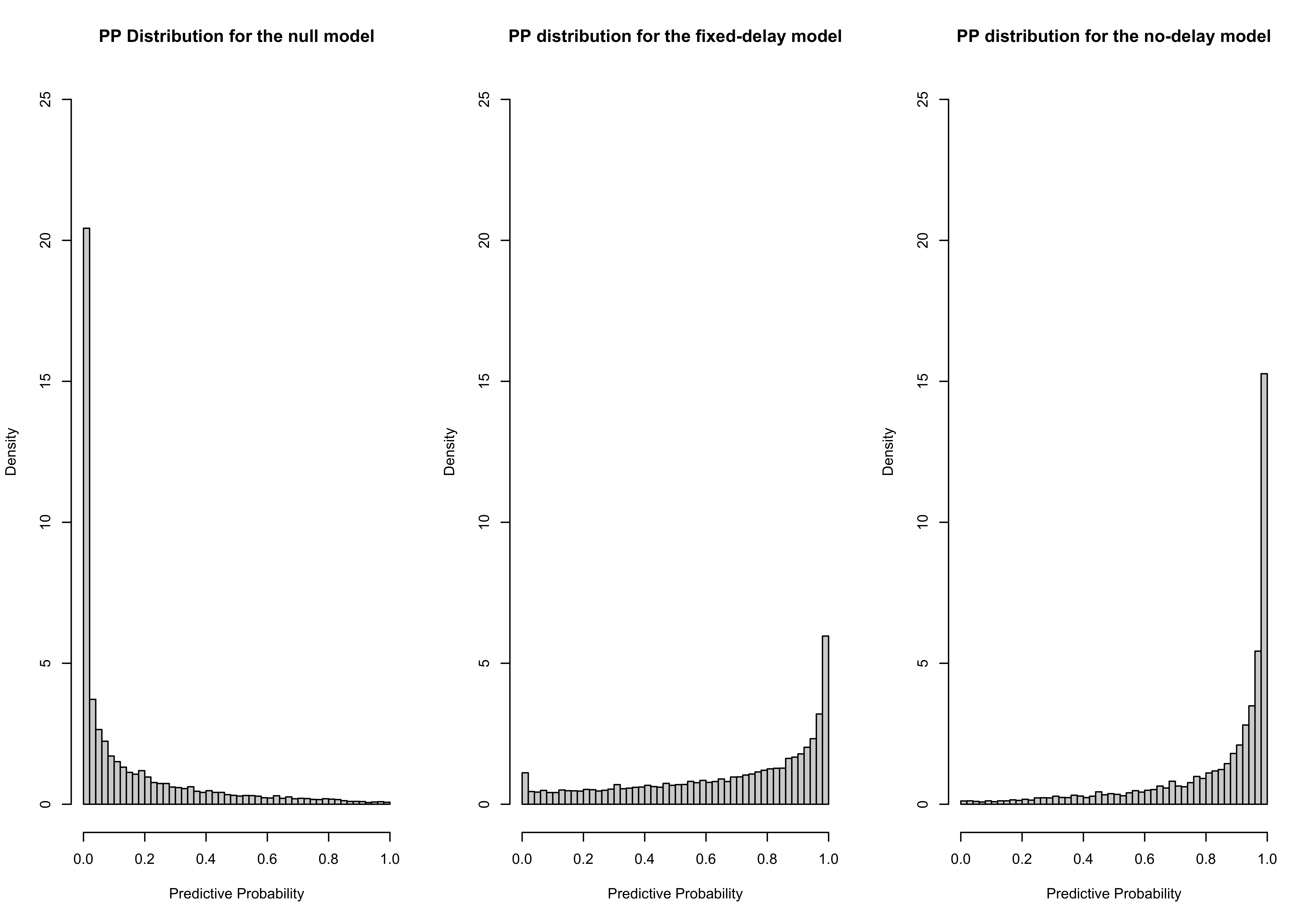}
\caption{PP distributions at the interim analysis (IF $= 0.50$) 
under the null, fixed-delay, and immediate-effect scenarios. The 
null distribution is concentrated near zero; both alternatives are 
skewed towards one, with the no-delay case more sharply 
concentrated than the fixed-delay case.}
\label{fig:PP_thresholds_hists}
\end{figure}

Figure~\ref{fig:PP_thresholds_hists} displays the PP distributions 
under each scenario. Under the null, the distribution is strongly 
concentrated near zero, with a sharp spike at low PP values and a 
long, thin tail towards one --- reflecting that interim data 
generated under no treatment effect are largely predictive of an 
unsuccessful final analysis. Under both alternatives, the 
distributions are skewed in the opposite direction, with mass 
concentrated near one. The no-delay distribution shows the 
sharpest concentration, with very little mass away from one, whilst 
the fixed-delay distribution is more diffuse, exhibiting a 
non-trivial left tail. This is expected: the delayed onset of the 
treatment effect means that at IF $= 0.50$, some of the 
treatment benefit has not yet fully manifested in the interim data, 
introducing additional uncertainty into the PP.

Table~\ref{tab:PP_thresholds} quantifies this separation across a 
range of candidate thresholds. Under the null, 
$\Pr(\mathrm{PP} < \lambda)$ is already $0.58$ at $\lambda = 0.05$, 
rising to $0.94$ at $\lambda = 0.50$, reflecting the strong 
concentration of null PP values near zero. Under both alternatives, 
the corresponding probabilities remain substantially lower across 
the full range. The no-delay scenario shows the smallest values 
throughout, whilst the fixed-delay scenario sits between the two, 
consistent with the additional uncertainty visible in 
Figure~\ref{fig:PP_thresholds_hists}. At $\lambda = 0.20$, for 
example, the probability of incorrectly triggering futility is 
$0.03$ under the no-delay alternative and $0.12$ under the 
fixed-delay alternative, compared with $0.81$ under the null --- 
a clear separation across all three scenarios.

\begin{table}[ht]
\centering
\caption{
Estimated $\Pr(\mathrm{PP}<\lambda)$ at the interim analysis (IF = 0.50) under each
scenario, based on 2{,}000 simulations.
}
\label{tab:PP_thresholds}
\small
\begin{tabular}{lccc}
\toprule
Threshold $\lambda$ & Null & Fixed-delay & No-delay \\
\midrule
0.05 & 0.5754 & 0.0316 & 0.0063 \\
0.10 & 0.6933 & 0.0568 & 0.0137 \\
0.15 & 0.7643 & 0.0864 & 0.0224 \\
0.20 & 0.8131 & 0.1159 & 0.0334 \\
0.25 & 0.8473 & 0.1435 & 0.0439 \\
0.30 & 0.8726 & 0.1737 & 0.0563 \\
0.35 & 0.8945 & 0.2028 & 0.0704 \\
0.40 & 0.9114 & 0.2347 & 0.0896 \\
0.45 & 0.9280 & 0.2664 & 0.1080 \\
0.50 & 0.9402 & 0.3031 & 0.1280 \\
\bottomrule
\end{tabular}
\end{table}

A threshold of $\lambda = 0.10$ was selected. At this value, the 
probability of correctly triggering futility under the null is 
$0.69$, whilst the probability of incorrectly stopping a trial 
with a delayed treatment effect remains low at $0.06$. This 
reflects a conservative calibration, prioritising protection 
against premature termination of trials with a genuine --- if 
delayed --- treatment effect, which is the primary motivation 
for the framework.

Overall, the separation of PP distributions across scenarios supports the use of
predictive probability as a futility criterion, even in the presence of delayed
treatment effects.

\subsubsection{Step 3: Evaluation of the Full Adaptive Design}

With the interim timing fixed at $\mathrm{IF}=0.50$ and the futility threshold set to $\lambda=0.10$, we evaluated the full adaptive design under four data-generating scenarios:

\begin{enumerate}
    \item \textbf{S1 (Null)}: no treatment effect.
    \item \textbf{S2 (Delayed alternative)}: 3-month delay followed by HR $=0.8$.
    \item \textbf{S3 (Immediate alternative)}: proportional hazards with HR = 0.8.
    \item \textbf{S4 (Prior predictive mixture)}: uncertainty integrated over the elicited prior distributions.
\end{enumerate}

These scenarios span both fixed effect assumptions (S1-S3) and the uncertainty structure implied by the elicitation (S4). In practice, a comprehensive design exercise would explore a broader range of hazard ratios, delay distributions, and recruitment patterns; however, the four scenarios considered here capture the principal operating regimes relevant to delayed-onset survival studies.

\vspace{0.4em}
\noindent
We compared four monitoring strategies:

\begin{itemize}
    \item \textbf{D1: Fixed design} with no interim monitoring.
    
    \item \textbf{D2: Group sequential design (GSD)} with a single efficacy look at $\mathrm{IF}=0.75$. 
    A one-sided $\alpha=0.025$ spending function allocates $0.0125$ at the interim look and $0.025$ at the final analysis, giving efficacy boundaries
    \[
    Z > 2.241 \quad (\text{interim}), \qquad
    Z > 2.047 \quad (\text{final}).
    \]
    
    \item \textbf{D3: Hybrid PP--GSD design}.  
    Uses the same efficacy boundaries as D2, supplemented with the predictive-probability futility rule as found in the previous sections
    \[
    \text{stop for futility at IF }=0.50 \text{ if } \textit{PP} < 0.10.
    \]
    
    \item \textbf{D4: $\alpha$- and $\beta$-spending GSD}.  
    The $\alpha$-spending matches D2. The futility boundary 
    was calibrated following the same procedure as D3: 
    2{,}000 interim datasets were simulated under the null, 
    fixed-delay, and immediate-effect scenarios, and the 
    log-rank $Z$ statistic was recorded at $\mathrm{IF}=0.50$. 
    To match the null 
    stopping probability of D3 ($\approx 69\%$), we selected
    \[
    \text{stop for futility at IF }=0.50 \text{ if } Z < 0.517 ,
    \]
    yielding a futility rate of approximately $14.7\%$ 
    under the fixed-delay alternative and $3.8\%$ under the 
    immediate-effect alternative.

\end{itemize}

For each design-scenario combination, we simulated 100{,}000 Monte Carlo replications to estimate:

\begin{itemize}
    \item type~I error (under S1), power (under S2-S3) or  assurance (under S4),
    \item probabilities of early stopping for efficacy and futility,  
    \item expected sample size, and  
    \item expected trial duration.  
\end{itemize}

\subsubsection{Results}

\begin{table*}[t]
\centering
\caption{Operating characteristics of the four monitoring strategies (D1: no interim analysis; D2: GSD efficacy-only; D3: GSD with predictive probability futility; D4: GSD with $\beta$-spending futility) under the four data-generating scenarios S1--S4. Estimates are based on 100{,}000 simulated trials.}
\label{tab:S1S3_D1D3}
\resizebox{\textwidth}{!}{
\begin{tabular}{llcccccc}
\toprule
\textbf{Scenario} & \textbf{Design} & \textbf{P(Reject $H_0$)} & \textbf{P(Early Fut.)} & \textbf{P(Early Eff.)} & \textbf{ESS} & \textbf{Duration} \\
\midrule
\textbf{S1: Null}
    & D1 & 0.0260 & -- & -- & 1200 & 29.47 \\
    & D2 & 0.0252 & -- & 0.0117 & 1200 & 29.40 \\
    & D3 & 0.0237 & 0.6866 & 0.0116 & 1023 & 22.12 \\
    & D4 & 0.0239 & 0.6959 & 0.0117 & 1022 & 22.03 \\

\midrule

\textbf{S2: 3-month delay, then HR = 0.8}
    & D1 & 0.7169 & -- & -- & 1200 & 30.63 \\
& D2 & 0.6948 & -- & 0.4162 & 1200 & 28.13 \\
& D3 & 0.6879 & 0.0550 & 0.4159 & 1187 & 27.52 \\
& D4 & 0.6623 & 0.1525 & 0.4146 & 1165 & 26.43 \\

\midrule

\textbf{S3: HR = 0.8}
    & D1 & 0.8961 & -- & -- & 1200 & 30.85 \\
& D2 & 0.8866 & -- & 0.7079 & 1200 & 26.61 \\
& D3 & 0.8831 & 0.0144 & 0.7077 & 1197 & 26.45 \\
& D4 & 0.8745 & 0.0386 & 0.7068 & 1191 & 26.19 \\

\midrule

\textbf{S4: Elicited Priors}
     & D1 & 0.6100 & -- & -- & 1200 & 30.73 \\
& D2 & 0.6008 & -- & 0.4732 & 1200 & 27.76 \\
& D3 & 0.5966 & 0.1957 & 0.4730 & 1150 & 25.66 \\
& D4 & 0.5861 & 0.2457 & 0.4723 & 1139 & 25.11 \\
\bottomrule
\end{tabular}
}
\end{table*}

Table~\ref{tab:S1S3_D1D3} summarises the operating characteristics of the four monitoring strategies across all four scenarios. We discuss each in turn before drawing overall conclusions.

\paragraph{Scenario S1}
Under the null (S1), all four designs control the one-sided Type I error rate at the nominal 2.5\% level, with estimated rejection probabilities ranging from 0.024 to 0.026. The efficacy-only GSD (D2) offers negligible efficiency gains over the fixed design (D1) in this setting, as early stopping for efficacy under the null is rare (1.2\%). By contrast, both adaptive designs (D3 and D4) stop approximately 69\% of null trials early for futility, reducing the expected sample size from 1,200 to approximately 1,022 patients and shortening mean trial duration by around 7 months. The two adaptive designs are nearly indistinguishable under S1 by construction: D4's futility boundary was explicitly calibrated to match D3's null stopping rate.

\paragraph{Scenario S2}
Under the delayed alternative (S2), the differences between D3 and D4 become apparent. This is the scenario of primary interest, as it represents the setting the framework is specifically designed to handle. D3's futility rule triggers in only 5.5\% of trials, which is consistent with the conservative calibration described in Section~\ref{sec:adaptive_example}, resulting in a power of 68.8\%, a loss of just 0.7 percentage points relative to D2. D4, despite matching D3's null stopping rate, triggers futility in 15.3\% of delayed-alternative trials, incurring a power loss of 3.3 percentage points relative to D2 and 0.9 percentage points relative to D1. This difference arises because the log-rank statistic at the interim, on which D4's futility boundary is based, is attenuated by the delayed onset of the treatment effect: at IF = 0.50, much of the survival benefit has yet to manifest, causing the $Z$-statistic to systematically underestimate the eventual treatment effect. The predictive probability, by contrast, incorporates prior information about the plausibility of a delayed effect and is therefore more robust to this attenuation. The result is that D3 approximately halves the power loss relative to D4 under the scenario of greatest clinical concern.

\paragraph{Scenario S3}
Under the immediate alternative (S3), both adaptive designs perform nearly identically and the differences from D2 are minimal. D3's futility rule fires in only 1.4\% of trials and D4's in 3.9\%, reflecting that a strong, immediate treatment effect produces interim data that are highly informative about eventual success. Power under D3 is 88.3\%, a loss of 0.4 percentage points relative to D2 and 1.3 relative to D1. The modest additional power loss of D4 relative to D3 (0.9 percentage points) is consistent with a small residual difference in futility stopping rates, but both designs preserve the vast majority of power available under proportional hazards.

\paragraph{Scenario S4}
Under the prior predictive scenario (S4), which integrates uncertainty over the full elicited prior distribution and therefore represents the most realistic operating condition, D3 achieves an assurance of 59.7\%, compared with 60.1\% for D2 and 61.0\% for D1. The 0.4 percentage point loss relative to D2 is accompanied by a 19.6\% probability of early futility stopping, a reduction in expected sample size of approximately 50 patients, and a saving of around 2 months in expected trial duration. D4 stops somewhat more frequently for futility (24.6\%), yielding a further modest reduction in expected sample size and duration, but at the cost of an additional 1.0 percentage point of assurance relative to D3. The prior predictive scenario thereby confirms the pattern observed under S2: D3 strikes a more favourable balance between early stopping efficiency and protection of power when delayed effects are a realistic possibility.

\paragraph{Discussion}
Overall, the results demonstrate that the hybrid PP--GSD design (D3) achieves meaningful efficiency gains over the fixed and efficacy-only designs under the null and prior predictive scenarios, whilst incurring only small reductions in power or assurance under the alternatives. Crucially, the comparison with D4 illustrates the practical value of the predictive probability approach: by incorporating prior information about the treatment effect structure, D3 is substantially more conservative under the delayed alternative than a standard beta-spending rule calibrated to the same null performance, at negligible additional cost in other scenarios. This asymmetry represents the central operating advantage of the proposed framework.

\subsection{Sensitivity Analysis}
We illustrate the sensitivity of the results to the prior probability parameters $P_S$ and $P_{\text{DTE}}$, following the procedure described in Section~\ref{sec:sens_analyses}. 

Table~\ref{tab:sensitivity} reports operating characteristics for the four designs under three alternative specifications of these parameters, with all other prior components held fixed.
The qualitative conclusions from S4 are robust to plausible changes in $P_S$ and $P_{\text{DTE}}$. Reducing $P_S$ from 0.9 to 0.8 lowers assurance across all designs, as expected, since a lower prior probability of any treatment effect reduces the overall probability of trial success. The relative ordering of the four designs is preserved throughout, and D3 consistently achieves a more favourable balance between assurance and early stopping efficiency than D4. Increasing $P_{\text{DTE}}$ from 0.7 to 0.8 has a smaller impact, reflecting that the primary driver of assurance in this example is whether a treatment effect exists at all, rather than whether it is delayed.

Further sensitivity analyses, including quartile reweighting of $D_{\text{HR}^*}$ and $D_T$, can be performed directly from the simulation outputs returned by \texttt{calc\_dte\_assurance\_adaptive()}, as described in Section~\ref{sec:sens_analyses}, though we do not illustrate these here.

\begin{table*}[t]
\centering
\caption{Operating characteristics of the four monitoring strategies (D1: no interim analysis; D2: GSD efficacy-only; D3: GSD with predictive probability futility; D4: GSD with $\beta$-spending futility) under alternative specifications of $P_S$ and 
$P_{\text{DTE}}$, with all other prior parameters held fixed. Estimates are based on 100{,}000 simulated trials.}
\label{tab:sensitivity}
\resizebox{\textwidth}{!}{
\begin{tabular}{lllcccccc}
\toprule
\textbf{$P_S$} & $P_{\text{DTE}}$ &  \textbf{Design} & \textbf{P(Reject $H_0$)} & \textbf{P(Early Fut.)} & \textbf{P(Early Eff.)} & \textbf{ESS} & \textbf{Duration} \\
\midrule
\textbf{0.8} & \textbf{0.7}
        & D1 & 0.5445 & -- & -- & 1200 & 30.59 \\
    & & D2 & 0.5363 & 0.4216 & -- & 1200 & 27.95 \\
    & & D3 & 0.5324 & 0.4214 & 0.2503 & 1136 & 25.27 \\
    & & D4 & 0.5231 & 0.4208 & 0.2964 & 1126 & 24.77 \\

\midrule

\textbf{0.9} & \textbf{0.8}
         & D1 & 0.6000 & -- & -- & 1200 & 30.71 \\
    & & D2 & 0.5902 & 0.4568 & -- & 1200 & 27.83 \\
    & & D3 & 0.5859 & 0.4566 & 0.2000 & 1149 & 25.69 \\
    & & D4 & 0.5743 & 0.4559 & 0.2535 & 1137 & 25.10 \\

\midrule

\textbf{0.8} & \textbf{0.8}
        & D1 & 0.5361 & -- & -- & 1200 & 30.57 \\
    & & D2 & 0.5274 & 0.4074 & -- & 1200 & 28.01 \\
    & & D3 & 0.5234 & 0.4072 & 0.2537 & 1135 & 25.30 \\
    & & D4 & 0.5131 & 0.4066 & 0.3030 & 1124 & 24.76 \\

\bottomrule
\end{tabular}
}
\end{table*}

\subsection{Software Implementation}

All computational procedures described in this section were implemented using the \texttt{DTEAssurance} R package developed as part of this research. The package provides a workflow that mirrors the three-step construction of the adaptive design:

\begin{itemize}
    \item \textbf{Timing calibration.} \\
   The function \texttt{calibrate\_PP\_timing()} automates prior-predictive simulation of interim datasets across candidate information fractions. For each candidate timing, it updates the posterior under the delayed-effect model, evaluates the predictive probability (PP), and summarises the discriminatory behaviour of the PP distribution. This supports principled selection of an information fraction at which the futility analysis is both informative and operationally meaningful. 

   \item \textbf{Threshold calibration.} \\
   The function \texttt{calibrate\_PP\_threshold()} evaluates the empirical distribution of the PP at the chosen interim time under multiple data-generating scenarios, including the null, fixed-delay alternatives, and immediate-effects models. These distributions guide selection of a futility threshold that balances protection against false-negative stopping with efficiency under unpromising scenarios.

   \item \textbf{Evaluation of the full adaptive design.} \\
   The function \texttt{calc\_dte\_assurance\_adaptive()} implements the complete hybrid monitoring strategy, combining the calibrated PP futility rule with the specified group-sequential efficacy boundaries. It simulates full trial trajectories, applies the interim and final decision rules, and returns design-level operating characteristics including power, type I error, early stopping probabilities, expected sample size, and expected trial duration.
   
\end{itemize}

The package is accompanied by an interactive \texttt{Shiny} application, which exposes the same computational workflow through a graphical interface. The app enables users to specify elicited priors, explore different delay structures, examine PP timing and threshold behaviour, and visualise operating characteristics for both fixed and adaptive designs. This facilitates communication with clinicians and trial statisticians, and provides a reproducible, accessible platform for practical design exploration.

\section{Discussion}\label{sec:discussion}

This paper has presented a principled and practical framework for 
designing adaptive clinical trials in the presence of delayed 
treatment effects, building directly on the elicitation 
methodology of Salsbury et al.\cite{Salsbury2024} By 
incorporating elicited prior distributions into both the 
assurance calculation and the interim decision-making process, 
the framework provides a coherent Bayesian approach to trial 
design that directly addresses the inferential and operational 
challenges that delayed treatment effects introduce.

\subsection{Summary of Contributions}

The first contribution of this paper is the extension of the 
assurance framework to adaptive designs with interim analyses. 
Assurance, the prior-weighted probability of trial success, 
provides a natural and interpretable summary of a trial's 
prospects under parameter uncertainty, and its extension to the 
adaptive setting allows trialists to evaluate how interim looks 
affect the overall probability of success. The simulation 
framework presented in section~\ref{sec:Framework} is general 
with respect to both the choice of test statistic and the 
futility stopping rule, allowing it to accommodate a wide range 
of adaptive designs.

The second and central contribution is the development of a 
Predictive Probability framework for futility 
decision-making at interim. By updating the elicited prior 
distributions with accumulating trial data via MCMC, and 
simulating forward the unobserved event process, the PP 
criterion directly targets the probability of ultimate trial 
success---even when most accumulated information lies in the 
pre-delay region. A key feature of the framework is that the 
delay time $T$ is treated as a free parameter that is updated 
with the data, rather than fixed at a pre-specified value. This 
reflects the genuine uncertainty about the onset of treatment 
benefit and avoids the sensitivity to misspecification that 
would arise from fixing $T$ arbitrarily. When coupled with 
conventional group sequential efficacy monitoring, this hybrid 
Bayesian-frequentist framework preserves strict Type I error 
control while enabling futility decisions that remain coherent 
under delayed or evolving treatment effects.

The third contribution is the calibration procedure for 
selecting the interim timing and PP threshold $\lambda$. By 
examining the distribution of PP values under the prior 
predictive distribution, the timing can be chosen to ensure the 
interim look is sufficiently informative before any data are 
collected. The threshold is then selected to achieve acceptable 
operating characteristics, including Type I error control, 
power under clinically meaningful scenarios, and minimisation 
of expected sample size under the null, providing a direct 
response to regulatory expectations around prespecification and 
calibration of adaptive designs.

The fourth contribution is the open-source \texttt{DTEAssurance} 
R package and accompanying Shiny application, which encapsulate 
the full workflow and provide a reproducible platform for design 
exploration. The Shiny application enables practitioners to 
visualise delay structures, interrogate predictive behaviour, 
and communicate design choices to clinical collaborators, 
lowering the barrier to adoption in applied settings.

\subsection{Practical Considerations}

While the statistical framework provides a principled basis for 
interim decision-making, we emphasise that futility stopping 
decisions in practice must also integrate clinical relevance, 
operational logistics, and ethical considerations. A low 
predictive probability at interim does not automatically 
mandate stopping, the decision should be made in the context 
of the full clinical picture by a suitably constituted Data 
Monitoring Committee.

We also note that the reliability of the posterior update 
depends on sufficient information having accrued by the time 
of the interim look. If the interim occurs before a meaningful 
proportion of patients have passed the delay time $T$, the 
likelihood will be largely uninformative about $\text{HR}^*$ 
and $T$, and the posterior will be heavily driven by the prior. 
The calibration procedure of Section~\ref{sec:calibration} 
directly addresses this by selecting the interim timing with 
reference to the prior distribution of $T$; nevertheless, 
trialists should be aware of this sensitivity, particularly 
when the prior on $T$ is diffuse.

\subsection{Limitations and Future Work}

Several limitations of the current framework warrant 
acknowledgement. First, the framework focuses on a single futility interim look. Whilst this is well-justified on both practical and computational grounds, each additional futility look requires a full MCMC fit and nested simulation, and extending to multiple looks would increase flexibility and is a natural direction for future work.

Second, the framework assumes a piecewise Weibull model with a 
common shape parameter. Whilst this is a flexible and 
well-motivated model for the delayed treatment effect setting, 
it may not capture all forms of non-proportional hazards that 
arise in practice. Extensions to more flexible survival models, for example, those based on splines or mixture 
distributions, are a natural direction for future work.

Third, the calibration procedure evaluates operating 
characteristics at a finite set of fixed point scenarios. 
Whilst these scenarios are chosen with reference to the 
elicited prior distributions, they do not provide a complete 
picture of the design's behaviour across the full parameter 
space. A more comprehensive assessment, for example, via 
Bayesian sensitivity analysis or global operating 
characteristic evaluation, would strengthen the calibration 
further.

Finally, whilst the framework is presented in the context of 
immuno-oncology, where delayed treatment effects are most 
commonly encountered, it has broad applicability to any 
therapeutic area in which a delayed biological mechanism is 
anticipated. We anticipate that the methodology will find 
application across a wide range of Phase III confirmatory 
trials.

\subsection*{ACKNOWLEDGEMENTS}
This work has been supported by a University of Sheffield EPSRC Doctoral Training Partnership (DTP) Case Conversion with Novartis Scholarship [project reference 2610753]. 

\subsection*{DATA AVAILABILITY STATEMENT}
Data sharing is not applicable to this article, as no new data were created or analysed in this study.

\bibliographystyle{unsrt}
\bibliography{references}

\appendix\label{sec:appendix}
\section{Supporting Software}
An R\cite{R2024} package, \texttt{\{DTEAssurance}\}, for implementing the methods described in this paper is available on GitHub, at \texttt{https://github.com/jamesalsbury/DTEAssurance}. The website also includes an illustration of using the package to replicate the examples in this paper. This package is installed with the commands
\\ \\
\texttt{install.packages("devtools")}

\noindent \texttt{devtools::install\char`_github("jamesalsbury/DTEAssurance")}.
\\ \\
    An app for implementing these methods, produced with \texttt{\{shiny}\},\cite{shiny} can be used online at \\ \texttt{ https://jamesalsbury.shinyapps.io/AdaptiveApp/}. 
A version of the app for offline use is included in the \texttt{DTEAssurance} package.

\section{Algorithms}

\begin{algorithm}[H]
\caption{Group Sequential Design with potential delayed treatment 
effect.}\label{alg:GSDDTE}

\textbf{Inputs:}
\begin{itemize}
    \item \textit{Trial design:} Sample sizes $n_c$, $n_e$; maximum 
    events $E \leq n_c + n_e$; $L$ planned interims with information 
    fractions $\mathbf{F} = (F_1, \ldots, F_L)$ and efficacy 
    boundaries $\mathbf{b} = (b_1, \ldots, b_{L+1})$; futility 
    rule $\mathcal{S}$ (Section~\ref{sec:PP}); test statistic 
    $Z(\cdot)$.
    \item \textit{Priors:} $\pi(\boldsymbol{\theta}_c \mid 
    \boldsymbol{x}_{\text{hist}})$; $D_{\text{HR}^*}$, $D_T$; 
    $P_S$, $P_{\text{DTE}}$.
    \item \textit{Simulation:} Iterations $N$; recruitment schedule.
\end{itemize}

\textbf{For} $i = 1, \ldots, N$, initialise $U_i = 0$ and:
\begin{enumerate}
    \item Sample $\boldsymbol{\theta}_{c,i} \sim 
    \pi(\boldsymbol{\theta}_c \mid \boldsymbol{x}_{\text{hist}})$ 
    and control survival times $x_{1,i}, \ldots, x_{n_c,i}$ from 
    the Weibull model (Equation~\ref{eq:control}).

    \item Sample $u \sim \mathrm{Uniform}(0,1)$. If $u < P_S$, 
    sample $\mathrm{HR}^*_i \sim D_{\mathrm{HR}^*}$ and, 
    independently, sample $T_i \sim D_T$ if $v \sim 
    \mathrm{Uniform}(0,1) < P_{\mathrm{DTE}}$, else set $T_i = 0$. 
    Otherwise set $\mathrm{HR}^*_i = 1$ and $T_i = 0$.

    \item Sample experimental survival times $y_{1,i}, \ldots, 
    y_{n_e,i}$ from the piecewise Weibull model with parameters 
    $\boldsymbol{\theta}_{c,i}$, $T_i$, $\mathrm{HR}^*_i$ 
    (Equation~\ref{eq:treatment}).

    \item Sample recruitment times $R_{1,i}, \ldots, 
    R_{(n_c+n_e),i}$ and compute pseudo-event times 
    $\mathcal{P}_{k,i} = R_{k,i} + t_{k,i}$.

    \item \textbf{For} $j = 1, \ldots, L$: let $E_{T,j}$ be the 
    calendar time of the $\lfloor F_j E \rfloor$-th event. Exclude 
    patients with $R_{k,i} > E_{T,j}$; censor those with 
    $\mathcal{P}_{k,i} > E_{T,j}$, setting their survival time to 
    $E_{T,j} - R_{k,i}$. Compute $Z_{i,j}$. If $\mathcal{S}$ 
    indicates futility, proceed to $i+1$. If $Z_{i,j} > b_j$, 
    set $U_i = 1$ and proceed to $i+1$.

    \item \textbf{Final analysis:} repeat the data construction of 
    step 5 at $E_{T,L+1}$ (the time of the $E$-th event). If 
    $Z_{i,L+1} > b_{L+1}$, set $U_i = 1$.
\end{enumerate}

\textbf{Output:}
\begin{equation*}
    \hat{A} = \frac{1}{N}\sum_{i=1}^{N} U_i.
\end{equation*}

\textit{Note:} $\mathcal{S}$ is treated as a general input; 
here it is defined via the predictive probability framework of 
Section~\ref{sec:PP}, though any pre-specified stopping rule 
may be substituted.
\end{algorithm}

\begin{algorithm}
\caption{Calculating predictive probability in the presence of a potential 
delayed treatment effect.}\label{alg:PPDTE}

\textbf{Inputs:} Maximum sample sizes $n_c$ and $n_e$; prior distributions 
$\pi(\boldsymbol{\theta}_c \mid \boldsymbol{x}_{\text{hist}})$, $\pi(T \mid S)$, 
and $\pi(\text{HR}^*)$; scenario probabilities $P_S$ and $P_{\text{DTE}}$; 
maximum number of events $E$ (with $E \leq n_c + n_e$); information fraction 
at the interim $L$; number of simulated trials $N$; number of inner iterations 
$M$.

\textbf{For} $i = 1, \ldots, N$:
\begin{enumerate}
    \item Sample $\boldsymbol{\theta}_{c,i} \sim \pi(\boldsymbol{\theta}_c \mid 
    \boldsymbol{x}_{\text{hist}})$.
    
    \item Sample control survival times $x_{1,i}, \ldots, x_{n_c, i}$ using 
    $\boldsymbol{\theta}_{c,i}$.
    
    \item Sample $u \sim \mathrm{Uniform}(0,1)$. If $u < P_S$, sample 
    $T_i \sim \pi(T \mid S)$ and $\mathrm{HR}^*_i \sim \pi(\mathrm{HR}^*)$; 
    otherwise set $T_i = 0$ and $\mathrm{HR}^*_i = 1$.
    
    \item Sample experimental survival times $y_{1,i}, \ldots, y_{n_e, i}$ 
    using $\boldsymbol{\theta}_{c,i}$, $T_i$, and $\mathrm{HR}^*_i$.
    
    \item Sample recruitment times $R_{1,i}, \ldots, R_{(n_c + n_e),i}$ from 
    the pre-specified recruitment schedule.
    
    \item Compute pseudo-event times $\mathcal{P}_{j,i} = R_{j,i} + t_{j,i}$ 
    for each subject $j$, where $t_{j,i}$ is the sampled survival time.
    
    \item Order $\{\mathcal{P}_{j,i}\}$ and define $E_{T,i}$ as the calendar 
    time at which the $\lfloor L \cdot E \rfloor$-th event is observed.
    
    \item Retain only subjects with $R_{j,i} \leq E_{T,i}$; censor any subject 
    for whom $\mathcal{P}_{j,i} > E_{T,i}$, redefining their observed survival 
    time as $E_{T,i} - R_{j,i}$.
    
    \item Update $\pi(\boldsymbol{\theta}_c \mid \boldsymbol{x}_{\text{hist}})$ 
    with the interim data to obtain the posterior 
    $\pi(\boldsymbol{\theta} \mid \boldsymbol{x}_{\text{int},i})$, via MCMC.
    
    \item \textbf{For} $j = 1, \ldots, M$:
\begin{enumerate}[label=(\roman*)]
        \item Sample $\boldsymbol{\theta}^{(j)}_i$ from 
        $\pi(\boldsymbol{\theta} \mid \boldsymbol{x}_{\text{int},i})$.
        \item Simulate the remaining unobserved data using 
        $\boldsymbol{\theta}^{(j)}_i$.
        \item Combine the observed interim data with the simulated data and 
        apply the pre-specified analysis method.
        \item Set $U_{i,j} = 1$ if the outcome is a success, and $U_{i,j} = 0$ 
        otherwise.
    \end{enumerate}
\end{enumerate}

The predictive probability for the $i$-th simulated trial is estimated as:
\begin{equation*}
    \widehat{\mathrm{PP}}_i = \frac{1}{M} \sum_{j=1}^{M} U_{i,j}.
\end{equation*}
\end{algorithm}

\end{document}